\documentclass[reprint,amsmath,amssymb,aps]{revtex4-2}

\usepackage{graphicx}% Include figure files
\usepackage{dcolumn}% Table formatting
\usepackage{xr}
\usepackage{hyperref}
\usepackage{marginnote}
\usepackage{xcolor}
\usepackage{soulutf8}
\newcommand{\Da}{\mathrm{Da}}
\newcommand{\DF}{\mathcal{D}_{F}}
\newcommand{\pos}{\boldsymbol{x}}
\newcommand{\posB}{\boldsymbol{w}}
\newcommand{\Rg}{R_{g}}
\newcommand{\Rc}{R_{c}}
\newcommand{\Nc}{N_{c}}
\newcommand{\dt}{\Delta t}
\newcommand{\normal}{\boldsymbol{n}}
\newcommand{\textda}{Damk\"{o}hler }
\newcommand{\lcut}{\sigma_{\mathrm{cut}}}
\newcommand{\posP}{\boldsymbol{X}}

\makeatletter
\newcommand*{\addFileDependency}[1]{% argument=file name and extension
  \typeout{(#1)}
  \@addtofilelist{#1}
  \IfFileExists{#1}{}{\typeout{No file #1.}}
}
\makeatother
\newcommand*{\myexternaldocument}[1]{%
    \externaldocument{#1}%
    \addFileDependency{#1.tex}%
    \addFileDependency{#1.aux}%
}
\myexternaldocument{supporting}

\newcommand*{\arxivmyexternaldocument}[2]{
 \externaldocument{#1/#2}%
   \addFileDependency{#2.tex}%
   \addFileDependency{#1/#2.aux}%
}
\arxivmyexternaldocument{build}{bsupporting}

\newcounter{mynote}
\newcommand*{\mynote}[2]{%
    #2%
	\refstepcounter{mynote}%
    \marginpar{\textcolor{blue}{\Alph{mynote}}}%
    \label{#1}%
}
\renewcommand*{\mynote}[2]{#2} % uncomment removes margin notes

\renewcommand{\color}[1]{} 

\begin{document}

\preprint{APS/123-QED}

\title{Continuum Limit of Dendritic Deposition}

\author{Daniel Jacobson}
\email{jacobson.daniel.r@gmail.com}
\affiliation{Division of Chemistry and Chemical Engineering, California Institute of Technology, Pasadena, California 91125, USA}
\author{Thomas F. Miller III}
\email{tom@iambic.ai}
\altaffiliation[Now at ]{Iambic Therapeutics, San Diego, CA, USA}
\affiliation{Division of Chemistry and Chemical Engineering, California Institute of Technology, Pasadena, California 91125, USA}

\date{\today}

\begin{abstract}
Continuum models are commonly used to study dendritic deposition in fields ranging from nonequilibrium statistical mechanics to battery research. However, the continuum approximation underlying these models is poorly understood, even in the simplified case of Brownian particles depositing onto a small, reactive cluster. Specifically, this system transitions from a compact to a dendritic morphology at a critical radius that depends on the particle size. But in simulations of the continuum (small-particle) limit, the critical radius does not reproduce the scaling predicted by a purely continuum analysis. \color{blue} This discrepancy suggests that continuum models may not be able to capture the microscopic physics of dendrite formation, \color{black} raising doubts about their experimental relevance. To clarify the continuum limit of dendritic deposition, here, we reexamine the critical radius scaling of the Brownian particle system using Brownian dynamics simulations. Compared to past studies, we probe larger system sizes, up to hundreds of millions of particles in some cases, and adopt an improved paradigm for the surface reaction. This paradigm allows us to converge our simulations and to work with well-defined physical parameters. Our results show that the critical radius scaling is, in fact, consistent with the continuum analysis, validating the continuum approach to modeling dendritic deposition. Nonetheless, the Brownian particle system converges to its continuum limit slowly. As a result, when applying continuum models to more complex deposition processes, the continuum approximation itself may be a significant source of error.
\end{abstract}

\maketitle

\section{Introduction}

Many systems that display dendritic growth consist of Brownian particles that deposit onto a reactive surface. Examples include colloidal aggregation, amyloid formation, and electroplating \cite{linUniversalityColloidAggregation1989, foderaElectrostaticsControlsFormation2013, kahandaColumnarGrowthKinetic1992}. Researchers in a number of fields, from nonequilibrium statistical mechanics to next-generation battery development, have long sought to understand this type of deposition \cite{sanderDiffusionlimitedAggregationKinetic2000, chazalvielElectrochemicalAspectsGeneration1990,liuMaterialsLithiumionBattery2018,linRevivingLithiumMetal2017, hwangSodiumionBatteriesPresent2017}. And one of the primary tools they use in pursuit of this goal is continuum modeling
\cite{wittenDiffusionLimitedAggregationKinetic1981,aogakiImageAnalysisMorphological1982,wittenDiffusionlimitedAggregation1983,chazalvielElectrochemicalAspectsGeneration1990,sundstromMorphologicalInstabilityElectrodeposition1995, sanderDiffusionlimitedAggregationKinetic2000,monroeEffectInterfacialDeformation2004, monroeImpactElasticDeformation2005,tikekarStabilityAnalysisElectrodeposition2014, liuStabilizingLithiumMetal2016,tuNanoporousHybridElectrolytes2017, choudhuryConfiningElectrodepositionMetals2018}.

Continuum models of dendritic deposition have been widely adopted because they offer three key advantages. First, they are flexible enough to account for a range of physical effects that influence the growth morphology, including interfacial stress and the electric field \cite{chazalvielElectrochemicalAspectsGeneration1990,monroeEffectInterfacialDeformation2004, monroeImpactElasticDeformation2005,tikekarStabilityAnalysisElectrodeposition2014}. Second, they can be simulated over large length and timescales \cite{chazalvielElectrochemicalAspectsGeneration1990,monroeDendriteGrowthLithium2003,liuStabilizingLithiumMetal2016}. And third, they are amenable to analytical methods. Specifically, a linear stability analysis of a boundary perturbation can often be used to identify the conditions that cause deposition to become dendritic as opposed to compact \cite{wittenDiffusionLimitedAggregationKinetic1981,wittenDiffusionlimitedAggregation1983,chazalvielElectrochemicalAspectsGeneration1990,sundstromMorphologicalInstabilityElectrodeposition1995, monroeEffectInterfacialDeformation2004,monroeImpactElasticDeformation2005, tikekarStabilityAnalysisElectrodeposition2014,tuNanoporousHybridElectrolytes2017, choudhuryConfiningElectrodepositionMetals2018}.

Yet, in spite of their ubiquity, the relationship between continuum models and the deposition processes they approximate is poorly understood. As a defining example of this point, consider one of the simplest forms of dendritic deposition: Brownian particles depositing from all sides onto a cluster that has a uniform reactivity \cite{wittenDiffusionlimitedAggregation1983}. Outside of their interaction with the cluster, the Brownian particles diffuse freely, and once they deposit, they are fixed in place, so there is no \mynote{marg:reverse_reaction}{reverse} reaction or surface relaxation \cite{kesslerPatternSelectionFingered1988}. A fundamental property of this system is that the thickness of the dendritic branches increases as the rate of the surface reaction decreases, a phenomenon that qualitatively matches electrodeposition experiments \cite{wittenDiffusionlimitedAggregation1983, kahandaColumnarGrowthKinetic1992}. However, even in this simple case, a continuum analysis of the branch thickness has not been found to be consistent with particle-based simulations \cite{wittenDiffusionlimitedAggregation1983,meakinModelsColloidalAggregation1988,halseyElectrodepositionDiffusionLimited1990}.

\begin{figure}
\includegraphics{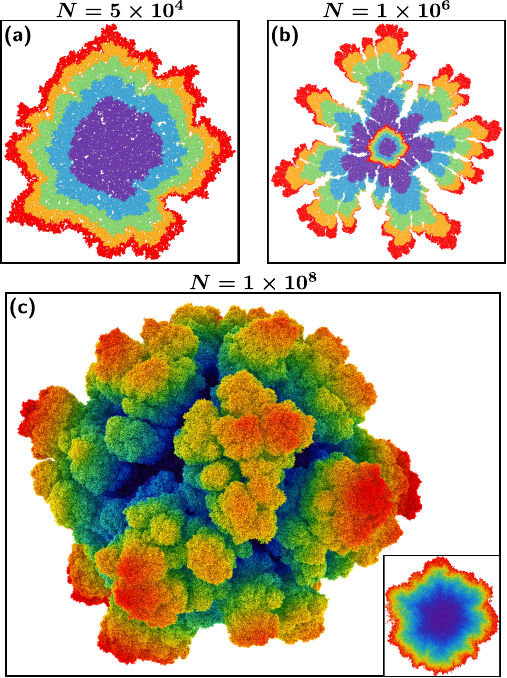}
    \caption{\label{fig:morphology} The compact-to-dendritic transition in the reactive deposition of Brownian particles at $\log_{10} \Da = -2.12$. Color schemes vary depending on the panel, as indicated in the following descriptions. Panels (a) and (b) depict two-dimensional deposition. The $N$ particles in these panels are rendered with a radius twice that of the actual radius for clarity. The $i$-th band of color moving outwards from the center corresponds to the structure after $iN / 5$ particles have been deposited. (a) Initially, the cluster displays a compact morphology. (b) However, upon reaching a critical radius, it spontaneously forms dendritic branches. (c) Main Panel: During three-dimensional deposition, a similar dendritic morphology emerges at large length scales. Particles are colored based on their distance from the origin. Inset: A two-dimensional slice through the initial compact 3D cluster ($N = 1 \times 10^{6}$). The color scheme follows a continuous gradient based on the deposition order.}
\end{figure}

In particular, rather than analyzing the branch thickness directly, it is more convenient to study the system's compact-to-dendritic (CTD) transition \cite{wittenDiffusionlimitedAggregation1983, meakinDiffusioncontrolledClusterFormation1983,meakinModelsColloidalAggregation1988, nagataniLaplacianGrowthPhenomena1989,halseyElectrodepositionDiffusionLimited1990}. This transition occurs when deposition is initialized from a small, compact cluster. While at first, the cluster continues to display a compact morphology as it grows, upon reaching a critical radius, it splits into a characteristic dendritic pattern, as shown in Fig.~\ref{fig:morphology} \footnotetext[1]{Meakin suggested that the initial phase of compact deposition is linked to Eden growth, reasoning that, when the cluster is small enough, incoming particles equilibrate with its surface and become equally likely to deposit everywhere \cite{meakinComputerSimulationGrowth1986,meakinFractalsScalingGrowth1998}.}\cite{Note1}. The size of the critical radius thus determines the branch thickness of the large-length-scale dendritic morphology.

A continuum linear stability analysis reveals that the critical radius for the CTD transition, $\Rc$, scales as \footnotetext[2]{See also Sec.~\ref{si:linear}.}\cite{wittenDiffusionlimitedAggregation1983, erbanReactiveBoundaryConditions2007, Note2}
\begin{equation}\label{continuum_scaling}
\Rc \sim D/k.
\end{equation}
Here, $D$ is the diffusion constant, and $k$ is the surface reaction rate constant. This scaling holds for both two- and three-dimensional deposition.

We can compare the prediction of \eqref{continuum_scaling} against the behavior of the particle-based deposition process. For this process, $\Rc$ depends on the dimensionless \textda number
\begin{equation}\label{Da_definition}
    \Da = \frac{k a}{D}, 
\end{equation}
where $a$ is the particle radius. Physically, $\Da$ quantifies the relative rate of reaction versus diffusion in the system. In the limit as the reaction becomes infinitely fast, $\Da \to \infty$, the process converges to diffusion-limited aggregation. Particles deposit as soon as they touch the cluster, and the CTD critical radius approaches zero, yielding single-particle thick dendritic branches \cite{wittenDiffusionLimitedAggregationKinetic1981,wittenDiffusionlimitedAggregation1983}.

The limit $\Da \to 0$, in contrast, corresponds to the continuum limit of the \textit{particle-based} process. That is, it can be reached by taking the particle radius $a$ to zero at fixed values of $k$ and $D$. Further, taking $\Da \to 0$ also yields a nontrivial critical radius $\Rc$ \cite{wittenDiffusionlimitedAggregation1983}. Consequently, it is this critical radius that might be expected to exhibit the scaling predicted by the continuum model. Dimensional analysis suggests that as $\Da \to 0$, $\Rc$ converges to the power-law form \footnotetext[3]{Choosing to nondimensionalize the critical radius by the particle radius $a$ provides a different, though equivalent, scaling relationship \cite{wittenDiffusionlimitedAggregation1983,meakinModelsColloidalAggregation1988, halseyElectrodepositionDiffusionLimited1990}.} \cite{barenblattScalingSelfsimilarityIntermediate1996, Note3}
\begin{equation}\label{critical_radius_scaling}
     \frac{k \Rc}{D} \sim \Da^{\gamma}.
\end{equation}
Here, $\gamma$ is an unknown scaling exponent that encodes the mechanism of the CTD transition. If \eqref{critical_radius_scaling} is to match the continuum analysis, \eqref{continuum_scaling}, then in two and three dimensions, $\gamma$ must be equal to zero \cite{wittenDiffusionlimitedAggregation1983}.

The $\gamma = 0$ hypothesis, however, has received limited empirical support. Refs.~\onlinecite{meakinModelsColloidalAggregation1988} and \onlinecite{halseyElectrodepositionDiffusionLimited1990} calculated $\gamma$ using lattice-based and off-lattice 2D Brownian dynamics simulations, obtaining estimates of $-0.2$ and $-0.25 \pm 0.3$, respectively \footnotetext[4]{See also Sections \ref{sec:methods} and \ref{si:lattice}.}\cite{Note4}. In the latter case, $\gamma = 0$ is included at the edge of the error bars, so at a minimum, this value cannot be ruled out. But taken together, these two studies suggest a power-law divergence with $\gamma \approx -0.2$.

The stakes for clarifying the discrepancy in the scaling behavior are high. If $\gamma$ is equal to zero, the continuum approximation works as intended. In this case, the many continuum models in the literature have a concrete theoretical foundation \cite{aogakiImageAnalysisMorphological1982,chazalvielElectrochemicalAspectsGeneration1990,sundstromMorphologicalInstabilityElectrodeposition1995, sanderDiffusionlimitedAggregationKinetic2000,monroeEffectInterfacialDeformation2004, monroeImpactElasticDeformation2005, tikekarStabilityAnalysisElectrodeposition2014, liuStabilizingLithiumMetal2016,tuNanoporousHybridElectrolytes2017, choudhuryConfiningElectrodepositionMetals2018}. Moreover, by mapping the rate of convergence to the $\gamma=0$ regime, we could then evaluate how much error results from applying the continuum approximation at a given \textda number. Quantifying this relationship in the Brownian particle system would be a step towards understanding how well continuum models work for more complex processes like electrodeposition.

\color{blue}

In contrast, if $\gamma \neq 0$, the particle-based process does not have a well-defined continuum limit. The critical radius either diverges for $\gamma < 0$ or collapses to zero for $\gamma > 0$. The absence of such a limit would be significant on its own terms. For example, by explaining this feature, we might be able to identify what general characteristics cause the continuum convergence of nonequilibrium systems to break down. Just as importantly, $\gamma \neq 0$ would also imply that the continuum analysis is fundamentally inaccurate. That is, it misses an essential part of the particle-based system's dendrite formation mechanism. 

This disconnect between the particle-based and the continuum views would be cause for concern. The inconsistency might result from a specific quirk of the particle-based system. However, it could also be a universal feature of dendritic deposition, one that persists even in the presence of additional physics, such as an electric field or surface relaxation. The inherent simplicity of the particle-based process makes this latter interpretation difficult to dismiss. Consequently, $\gamma \neq 0$ would cast doubt upon the reliability of continuum dendritic deposition models as a whole \cite{aogakiImageAnalysisMorphological1982,chazalvielElectrochemicalAspectsGeneration1990,sundstromMorphologicalInstabilityElectrodeposition1995, sanderDiffusionlimitedAggregationKinetic2000,monroeEffectInterfacialDeformation2004, monroeImpactElasticDeformation2005, tikekarStabilityAnalysisElectrodeposition2014, liuStabilizingLithiumMetal2016,tuNanoporousHybridElectrolytes2017, choudhuryConfiningElectrodepositionMetals2018}.

\color{black}

Given that determining the value of $\gamma$ is critical for understanding the continuum limit of dendritic deposition, here, we reexamine this exponent using an updated Brownian dynamics simulation approach. We improve on past simulation studies in four ways. First, we use a corrected simulation algorithm \cite{erbanReactiveBoundaryConditions2007, singerPartiallyReflectedDiffusion2008}. While previous authors fixed the probability that the particle deposits onto the cluster upon contact, this procedure makes achieving timestep convergence impossible \cite{halseyElectrodepositionDiffusionLimited1990, mayersSuppressionDendriteFormation2012}. Instead, the sticking probability must be a function of the timestep. Second, we carefully evaluate the convergence of $\gamma$ as the \textda number is taken to zero. Third, we examine the CTD transition in two dimensions at much smaller values of the \textda number than those used to probe the continuum limit in previous work \cite{meakinModelsColloidalAggregation1988, halseyElectrodepositionDiffusionLimited1990}. As part of this process, we generate clusters containing hundreds of millions of particles, orders of magnitude larger than the largest cluster sizes reported so far. Finally, we also investigate, for the first time, the scaling of the CTD transition in three dimensions.

Apart from our simulation-based analysis of $\gamma$, we also review the literature relating to the CTD transition. In particular, we reorganize this literature into a unified framework based on the \textda number introduced in \eqref{Da_definition}. Previously, the sticking probability in the Brownian dynamics algorithm was used in place of $\Da$ because it was taken to be a fundamental physical parameter \cite{nagataniLaplacianGrowthPhenomena1989,nagataniDoublecrossoverPhenomenaLaplacian1990,halseyElectrodepositionDiffusionLimited1990, meakinFractalsScalingGrowth1998,mayersSuppressionDendriteFormation2012}. However, the sticking probability is actually a convergence parameter akin to the simulation timestep \cite{erbanReactiveBoundaryConditions2007, singerPartiallyReflectedDiffusion2008}.

The rest of the manuscript is organized as follows. In Sec.~\ref{sec:model}, we define the particle-based deposition process and the associated continuum model. Next, in Sec.~\ref{sec:methods}, we describe our Brownian dynamics algorithm for simulating particle-based deposition and detail our methods for calculating the exponent $\gamma$. Having covered these preliminaries, in Sec.~\ref{sec:results}, we evaluate $\gamma$ based on simulation data and discuss how our results compare to the predicted continuum scaling, $\gamma = 0$. We conclude in Sec.~\ref{sec:conclusion} by addressing the implications of our findings for the study of more complex dendritic deposition processes.

\section{Model Definitions}\label{sec:model}

In this section, we first describe the Brownian particle deposition process and then show how this system is modeled at the continuum scale. 

The particle-based deposition process proceeds as follows. Initially, the cluster is composed of a single reactive particle with radius $a$, fixed at the origin \cite{wittenDiffusionlimitedAggregation1983}. A Brownian particle, also with radius $a$, is launched from a random point on a circle (or a sphere in 3D) with radius $b$ that surrounds the cluster. After this particle deposits, another Brownian particle is launched, and the cycle repeats. The launching surface is destructive. If the Brownian particle ever returns to this surface, it is killed, and a new particle is introduced. Lastly, the launching radius $b$ is assumed to be very large. That is, we take the limit $b \to \infty$ (details of how we handle this limit in simulations can be found in Sec. \ref{si:brownian}).

Before making the dynamics more precise, we simplify the excluded volume interaction. Specifically, we treat the incoming particle as a point particle and double the radii of the cluster particles, generating what we term the ``supercluster."

The key quantity that defines the deposition process is the growth probability density, $\rho(\posB)$, the probability density that the incoming particle attaches to the supercluster boundary at the point $\posB$. $\rho(\posB)$ can be expressed in terms of a steady-state concentration field $C$ using the standard Laplacian growth framework \cite{niemeyerFractalDimensionDielectric1984, pietroneroStochasticModelDielectric1984}. Following this approach, the field $C(\pos)$ (for position $\pos$) satisfies Laplace's equation 
\begin{equation}\label{laplace_equation}
    \nabla^2 C = 0
\end{equation}
on the domain outside of the supercluster boundary and inside the launching radius. The launching circle or sphere becomes a particle bath with an arbitrary, fixed concentration $C_{0}$
\begin{equation}\label{bath_bc}
    C(|\pos| = b) = C_0.
\end{equation}
And lastly, the reactivity of the supercluster at a point $\posB$ is included with the boundary \mynote{marg:cusps}{condition} \footnotetext[7]{The outward unit normal $\normal$ is not well-defined at the cusp points on the supercluster boundary where particles intersect. We can construct a formal interpretation of these cusps by taking them to be smooth over a scale $\aleph$. We then examine the growth behavior of the system in the limit they become sharp, $\aleph \to 0$. In our simulations, cusps are actually handled with rejection moves in a manner that is roughly akin to having $\aleph / a = 10^{-8}$. See the $\iota$ parameter in Sec.~\ref{si:brownian} for further details.}\cite{Note7}
\begin{equation}\label{reactive_bc}
	D \nabla C(\posB) \cdot \normal(\posB) = k C(\posB).
\end{equation}
Here, $\normal$ is the outward unit normal, $D$ is the diffusivity of the incoming particle, and $k$ is the reaction rate constant. Note that $k$ is a surface rate constant, and so has units of velocity. The growth probability density $\rho(\posB)$ is then proportional to the flux
\begin{equation}\label{growth_probability_density}
    \rho(\posB) \propto \nabla C(\posB) \cdot \normal(\posB).
\end{equation}

The equations that govern the deposition dynamics, \eqref{laplace_equation} through \eqref{growth_probability_density}, contain three dimensional parameters $k$, $a$, and $D$. These parameters combine to form the \textda number, $\Da$ in \eqref{Da_definition}, which sets the ratio of the reaction and diffusion rates in the system. In addition, since it is the only dimensionless quantity, $\Da$ uniquely determines the growth morphology.

We now show how the Brownian particle deposition process is course-grained to form its continuum model with two changes. First, the incoming particles are represented with a concentration field $C$ \cite{wittenDiffusionlimitedAggregation1983}. We use the same label $C$ for this field as we use for the Laplacian growth field for reasons that will become clear shortly. Second, the reactive cluster becomes a smooth closed curve (or a surface in 3D), as opposed to a collection of deposited particles. Just as in the microscopic process, this cluster evolves over time. However, rather than expanding particle-by-particle, its boundary grows outwards at every point simultaneously with a velocity $v(\posB)$ that depends on the flux
\begin{equation}\label{velocity}
    v(\posB) =  \mu D \nabla C(\posB) \cdot \normal(\posB).
\end{equation}
The constant of proportionality in \eqref{velocity}, $\mu$, is taken to be small enough that $C$ is pseudosteady. Consequently, this concentration field is described by the same set of equations, \eqref{laplace_equation}-\eqref{reactive_bc}, as the Laplacian growth field in the microscopic dynamics. 

\section{Theory and Methods}\label{sec:methods}

\subsection{Brownian Dynamics Algorithm}

To propagate the dynamics of the particle-based deposition process, we add a particle to the cluster at a point drawn from the growth probability density $\rho$ defined in \eqref{growth_probability_density}. 
We obtain this $\rho$  sample by simulating the motion of the incoming particle directly. Specifically, we use the specialized Brownian dynamics algorithm from Refs.~\onlinecite{erbanReactiveBoundaryConditions2007} and \onlinecite{singerPartiallyReflectedDiffusion2008}.  In this subsection, we provide an overview of this algorithm. Further implementation details can be found in Sec.~\ref{si:brownian}. Our code is also publicly available \cite{zenodo1}.

The Brownian dynamics algorithm works in the following way. Each timestep, the position $\posP$ of the incoming particle is updated according \mynote{marg:gaussian_update}{to}
\begin{equation}\label{gaussian_update}
    \posP(t + \dt) = \posP(t) + \sqrt{2 D \dt} \boldsymbol{g}
\end{equation}
where $\boldsymbol{g}$ is a $d$ dimensional Gaussian random variable with zero mean and unit variance \cite{allenComputerSimulationLiquids2017}. If the particle makes contact with the cluster during the update, it deposits with a sticking probability $P$ \cite{erbanReactiveBoundaryConditions2007, singerPartiallyReflectedDiffusion2008}. Otherwise, it reflects off of the cluster surface. 

\mynote{marg:algorithm_summary}{The} algorithm, as described, is sufficient for integrating the dynamics, but is computationally inefficient. The goal of the simulation is to build up a large cluster as quickly as possible so that we can analyze the critical radius. However, if the incoming particle drifts far from the cluster, it will take many Gaussian steps to return to a position where it can potentially stick again.

The standard solution to this efficiency problem is to split the dynamics into a near and a far regime \cite{sanderFractalsFractalCorrelations1994, tolmanOfflatticeHypercubiclatticeModels1989}. We implement such a split by capping the maximum distance the particle can jump per Gaussian timestep at $\lcut$. We then use this cutoff distance to build a cell list that allows us to quickly determine if the incoming particle is within one cell of the cluster \cite{allenComputerSimulationLiquids2017}. If it is, we take a Gaussian step. Otherwise, we take a much larger step. In particular, we first use a $k$-d tree to find the minimum distance $H$ between the particle and the cluster \cite{tolmanOfflatticeHypercubiclatticeModels1989}. We then take a step in a random direction with a magnitude slightly less than $H$ (implicitly avoiding the possibility of a cluster-particle collision). In addition, when the particle drifts outside a circle or sphere that bounds the cluster, we return it to that bounding surface by integrating the dynamics analytically \cite{sanderFractalsFractalCorrelations1994}.

The $k$-d tree steps and the bounding surface steps are exact. Consequently, the accuracy of the dynamics is limited only by the Gaussian steps. Since the convergence parameters for these steps are the cutoff distance and timestep, any observable of interest in the simulation must be evaluated in the joint limit $\lcut \to \infty$ and $\dt \to 0$.

We now return to the part of the algorithm that resolves collisions generated during Gaussian steps \cite{erbanReactiveBoundaryConditions2007, singerPartiallyReflectedDiffusion2008}. When the particle contacts the cluster, it sticks with probability $P$. Otherwise, it bounces off. This process is unusual for two reasons. First, we would like to simulate deposition at a given value of the rate constant $k$. However, $k$ does not appear in the operational parameters of the algorithm, which include the particle radius $a$, the diffusion constant $D$, the timestep $\dt$, and the sticking probability $P$. Rather, the value of $k$ is set implicitly by the values of these other parameters. The second reason the algorithm is unusual is that the implicit equation for $k$ includes the timestep. Specifically, we have \cite{erbanReactiveBoundaryConditions2007, singerPartiallyReflectedDiffusion2008}
\begin{equation}\label{k_timestep}
    k = P \sqrt{\frac{D}{\pi \dt}}.
\end{equation}

Eq.~\eqref{k_timestep} implies that we must approach timestep convergence carefully. We still need to take the limit $\dt \to 0$ to converge any observables of interest calculated from the simulations. But if we take this limit for a fixed value of the sticking probability $P$, it will cause $k$ to diverge. Instead, to keep $k$ constant, we have to take the timestep and the sticking probability to zero simultaneously such that the ratio $P / \sqrt{\dt}$ remains fixed. In other words, to simulate a given $k$ value, we must always set
\begin{equation}\label{sticking_probability}
    P = k \sqrt{\frac{\pi \dt}{D}}
\end{equation}
as we take $\dt \to 0$. For this reason, the sticking probability is effectively a convergence parameter like the timestep rather than a physical parameter like the rate constant.

We verified the timestep and cutoff distance convergence of our simulations by evaluating the critical radius (defined in Sec.~\ref{subsec:critical_radius}) at various values of the \textda number. More information about this procedure can be found in Sec.~\ref{si:brownian}.

\subsection{Toy Model: A Brownian Particle in a One-Dimensional Box with Reactive Walls}

Since the convergence behavior of the Brownian dynamics algorithm is counterintuitive, we illustrate it here with a toy example. Consider a Brownian point particle initialized at a random position in a one-dimensional box of length $L$. The left wall of the box, wall $A$, is reactive with surface rate constant $k_{A}$, while the right wall of the box, wall $B$, is absorbing (or equivalently $k_{B} = \infty$). After nondimensionalizing with $L$ and the diffusion constant $D$, the probability density of the particle $\xi$ as a function of position $y$ and time $\tau$ is described by
\begin{equation} \label{toy_model}
\begin{gathered}
    \partial_{\tau} \xi =  \partial_{yy} \xi, \\
    \partial_{y} \xi(0, \tau) = \psi_{A} \xi(0, \tau), \\
    \xi(1, \tau) = 0, \\
    \xi(y, 0) = 1. \\
\end{gathered}
\end{equation}
Here, $\psi_{A} = k_{A}L/D$ is a dimensionless rate constant similar to the \textda number in the particle-based deposition process.

For this toy system, we focus on the effect of the timestep on the probability that the particle will react with wall $A$ on the left, $\Phi_A$. This quantity can be calculated analytically as follows. First, we integrate \eqref{toy_model} and define $\Xi = \int_0^{\infty} \xi d \tau$ yielding
\begin{equation}
\begin{gathered}
  	\Xi''(y) = -1  \\
   	\Xi'(0) = \psi_{A} \Xi(0), \\
    \Xi(1) = 0.
\end{gathered}
\end{equation}
$\Phi_A$ is then equal to the integrated flux
\begin{equation}\label{Phi_A}
	\Phi_{A} = \int_{0}^{\infty} \partial_{y}\xi(0, \tau) d\tau = \Xi'(0) = \frac{\psi_{A} / 2}{1 + \psi_{A}}.
\end{equation}
When the rate constant $\psi_A$ goes to infinity, $\Phi_A$ approaches $ 1/2$ as expected from symmetry.

\begin{figure}
  \includegraphics{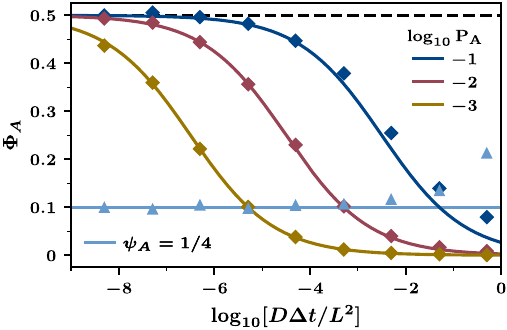}
  \caption{\label{fig:algorithm_demo_figure} The probability that the Brownian particle reacts with wall $A$, $\Phi_A$, for various values of the timestep $\dt$. Light blue triangles correspond to simulations run at $\psi_{A} = 1/4$ where the sticking probability at wall $A$, $P_A$, varies according to \eqref{sticking_probability}. As $\dt \to 0$, these simulations approach the analytical result from \eqref{Phi_A}, $\Phi_A = 1/10$ (light blue horizontal line). In contrast, the dark blue, violet, and brown diamonds were generated from simulations that used the fixed values of the sticking probability indicated in the upper-right legend. As $\dt \to 0$, these data approach $\Phi_A = 1/2$ (black-dashed line), the infinitely fast reaction limit, following the predictions of \eqref{k_timestep} and \eqref{Phi_A} (dark blue, violet, and brown curves). All numerical data points were generated using $10^4$ samples. Error bars constructed from Wald confidence intervals are smaller than the symbols \cite{agrestiApproximateBetterExact1998}.}
\end{figure}

In Fig.~\ref{fig:algorithm_demo_figure}, we demonstrate the convergence behavior of the Brownian dynamics algorithm by comparing $\Phi_A$ in simulations run with variable and fixed values of $P_A$, the sticking probability at wall $A$. This figure shows that only a variable sticking probability, \eqref{sticking_probability}, is consistent with simulating a fixed rate constant in reactive deposition. As $\dt \to 0$, the simulations run with this method at $\psi_{A} = 1/4$ (light blue triangles) approach the analytical solution $\Phi_A = 1/10$ from \eqref{Phi_A} (light blue line). In contrast, as predicted by \eqref{k_timestep}, the fixed sticking probability simulations (dark blue, violet, and brown diamonds) all converge to the infinitely fast reaction result, $1/2$, (black-dashed line).

\subsection{Previous Studies Treated the Sticking Probability as a Physical Parameter}

We now compare our conceptual framework for the Brownian particle deposition process, which is based on the \textda number, with the approach used by previous authors, which is based on the sticking probability. In particular, we focus on off-lattice deposition, where the particles move in a continuous space. Many authors instead studied lattice-based systems, which we discuss in Sec.~\ref{si:lattice} \cite{wittenDiffusionlimitedAggregation1983, meakinDiffusioncontrolledClusterFormation1983,meakinDiffusioncontrolledDepositionFibers1983,meakinModelsColloidalAggregation1988}.

Prior studies considered the sticking probability $P$ to be a fundamental physical parameter of the system that controlled the surface reaction rate rather than a convergence parameter like $\dt$ \cite{halseyElectrodepositionDiffusionLimited1990,nagataniLaplacianGrowthPhenomena1989,nagataniDoublecrossoverPhenomenaLaplacian1990,mayersSuppressionDendriteFormation2012}. This view of the sticking probability appeared to be borne out in simulations where different fixed values of $P$ were found to generate different cluster morphologies \cite{halseyElectrodepositionDiffusionLimited1990, mayersSuppressionDendriteFormation2012}. Consequently, Halsey and Leibig proposed that the critical radius for the CTD transition should scale as a power law with $P$ in the limit $P \to 0$. They then attempted to use simulations to calculate the associated power-law exponent \cite{halseyElectrodepositionDiffusionLimited1990}.

Eq.~\eqref{k_timestep} and Fig.~\ref{fig:algorithm_demo_figure}, however, illustrate that treating the sticking probability as a normal physical parameter has several conceptual limitations. In particular, all fixed values of $P$ yield the same infinitely fast reaction dynamics in the limit $\dt \to 0$. The reason different $P$ values previously seemed to produce different cluster morphologies in simulations was due to incomplete timestep convergence \cite{halseyElectrodepositionDiffusionLimited1990,mayersSuppressionDendriteFormation2012}. Moreover, since for any constant $P$, the morphology always crosses over immediately to fractal growth, examining the behavior of the critical radius in the limit as $P \to 0$ is not physically meaningful. 

To resolve the issues with the sticking probability, we can recast the simulation results of prior studies into the rate constant-based framework we have introduced here. First, we note that since these studies happened to use the same timestep for all of their simulations, the effective rate constant $k$ being simulated, \eqref{k_timestep}, is always directly proportional to the sticking probability $P$ \cite{halseyElectrodepositionDiffusionLimited1990, mayersSuppressionDendriteFormation2012}. As a result, the figures and calculations in these references can be adapted to our framework if the appropriate value of the \textda number is substituted for each value of $P$. By following this procedure, we find Halsey and Leibig's estimate of the power-law scaling exponent of the critical radius in terms of the sticking probability can instead be taken as an estimate for the exponent $\gamma$ in \eqref{critical_radius_scaling}, yielding $\gamma = -0.25 \pm 0.3$ \footnotetext[5]{Ref.~\onlinecite{meakinFractalsScalingGrowth1998} states that Halsey and Leibig's result in Ref.~\onlinecite{halseyElectrodepositionDiffusionLimited1990} is consistent with $\gamma = 0.2$ instead of $\gamma = -0.25$. However, we believe this is in error based on Eq.~$3.4$ in the original reference.}\cite{halseyElectrodepositionDiffusionLimited1990,Note5}.

There is, however, one potential pitfall that arises when converting from $P$ to $\Da$. After the swap, the simulation results cannot be assumed to be converged with respect to the timestep. That is, the only way to guarantee $\dt$ convergence at finite rate constant is to follow the prescription of \eqref{k_timestep} and take $\dt$ and $P$ to zero simultaneously. But previous authors instead always treated $P$ as a fixed quantity \cite{halseyElectrodepositionDiffusionLimited1990, mayersSuppressionDendriteFormation2012}. Nevertheless, the $\dt$ values these authors happened to choose are comparable to the $\dt$ value we selected in this work based on a rigorous convergence procedure (see Sec.~\ref{si:brownian}). Consequently, these studies' $P$ to $\Da$ converted results are likely free of significant timestep-related artifacts.

\subsection{The Reaction-Diffusion Length}\label{sec:rdl}

The ratio of the diffusion constant to the rate constant, $D / k$, plays a significant role in the Brownian particle system's CTD transition. As a reminder, in the continuum deposition model, we have $\Rc \sim D/k$ directly, \eqref{continuum_scaling}, and for the particle-based process, we have $\Rc \sim D/k$ in the continuum limit when $\gamma = 0$, \eqref{critical_radius_scaling}. \color{blue} From a macroscopic perspective, $D / k$ is the length over which the surface concentration on the reactive cluster is uniform \cite{sapovalGeneralFormulationLaplacian1994, filocheCanOneHear1999, grebenkovMathematicalBasisGeneral2006}. We show here how the Brownian dynamics algorithm helps provide an equivalent microscopic interpretation of this length scale.

\color{black}

According to the algorithm, the ratio of the sticking probability and the square root of the timestep is one of the key physical parameters in the simulation. This ratio defines a new microscopic reaction \mynote{marg:kappa_parameter}{parameter} $\kappa$
\begin{equation}\label{micro_to_macro}
    \kappa = \frac{P}{\sqrt{\dt}} = k \sqrt{\frac{\pi}{D}}
\end{equation}
that can be used in place of the macroscopic rate constant $k$ \cite{erbanReactiveBoundaryConditions2007, singerPartiallyReflectedDiffusion2008}. 

Thinking in terms of $\kappa$ is helpful for understanding the physical significance of $D / k$. Specifically, since it has units of reciprocal square root of time, $\kappa$ implies that the timescale for the surface reaction $T$ is
\begin{equation}\label{reaction_timescale}
    T \sim 1 / \kappa^{2} \sim D / k^{2}.
\end{equation}
The definition of $T$ is not immediately apparent based on the macroscopic view of the system since, if we start from the macroscopic rate constant $k$, we find both $a / k$ and $D / k^{2}$ have units of time. \color{blue} Based on \eqref{reaction_timescale}, we can see that the ratio $D/k$ is the length a particle can diffuse in the characteristic reaction timescale \cite{sapovalGeneralFormulationLaplacian1994, filocheCanOneHear1999}
\begin{equation}
    \sqrt{D T} \sim D / k.
\end{equation}
Consequently, we call this quantity the ``reaction-diffusion length.'' 

\color{black}

The reaction-diffusion length offers a new interpretation of the CTD transition's $\Rc \sim D/k$ scaling. Specifically, this scaling implies the CTD transition initiates when particles can no longer diffuse around the circumference of the cluster within the characteristic reaction timescale. If $\gamma = 0$, this mechanism helps explain dendrite formation in both the Brownian particle system and its continuum model. However, if $\gamma \neq 0$, the CTD transition of the Brownian particle system must result from some other physical process, one that the continuum perspective fails to adequately capture.

\subsection{The Critical Radius $\Rc$}\label{subsec:critical_radius}

\begin{figure*}
    \includegraphics{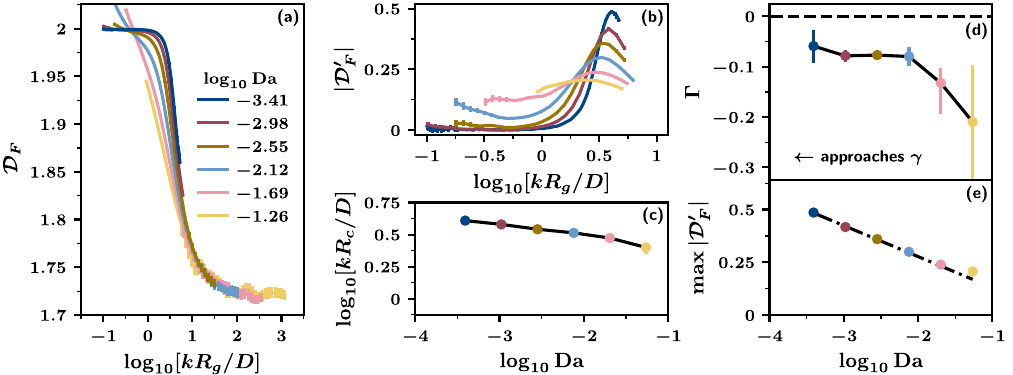}
    \caption{\label{fig:scaling} Continuum limit scaling of the compact-to-dendritic transition in two dimensions. All error bars are bootstrapped $95\%$ confidence intervals (CI) and are sometimes smaller than the lines or symbols. See Table \ref{tab:scaling} for further trajectory details. (a) The fractal dimension $\DF$ versus the radius of gyration $\Rg$ for various values of the \textda number, $\Da$. (b) Absolute value of the derivative of the fractal dimension $|\DF'|$. The peaks define the critical radius at each value of $\Da$, see \eqref{critical_radius}. (c) The critical radius $\Rc$ as a function of \textda number. (d) The derivative of panel (c), $\Gamma$, which provides an estimate for the scaling exponent $\gamma$ in \eqref{critical_radius_scaling} as $\Da \to 0$. The leftmost (dark blue) point has a value of $-0.06$ with a $95\%$CI of $[-0.03, -0.09]$. The black dashed line at the top of the panel represents the linear stability analysis prediction that $\gamma = 0$. (e) The maximum value of $|\DF'|$ (or equivalently $|\DF'|$ evaluated at critical radius) versus the \textda number. A linear fit to the three points with the lowest $\Da$ values (dash-dot black line) has a slope of $-0.15$ with a $95\%$CI of $[-0.16, -0.14]$.}
\end{figure*}

In this subsection, we define the critical radius quantitatively so that we can calculate it in simulations. The definition we use is posed in terms of the instantaneous fractal dimension of the cluster
\begin{equation}\label{fractal_dimension}
    \DF = \frac{d \log_{10} N}{d \log_{10} (k \Rg / D)}.
\end{equation}
Here, $N$ is the number of particles in the cluster, and $\Rg$ is the cluster's radius of gyration. For compact growth, $\DF$ approaches the dimension of the space $d$ after initial transients decay. In contrast, for dendritic growth, $\DF$ plateaus at a characteristic value less than $d$, once again, after initial transient behavior. The value of $\DF$ for 2D deposition in the dendritic regime has empirically been found to be $1.71$ independent of the value of $\Da$ \cite{meakinDiffusioncontrolledClusterFormation1983,meakinModelsColloidalAggregation1988, halseyElectrodepositionDiffusionLimited1990}. Whether the \textda number affects the $\DF$ for dendritic growth in 3D has not been explored, but in the limit $\Da \to \infty$, $\DF = 2.51$ \cite{tolmanOfflatticeHypercubiclatticeModels1989}. As a result of the two plateaus for the different morphologies, plots of the fractal dimension versus the log of the cluster radius during the CTD transition, such as Fig.~\ref{fig:scaling}(a), are roughly sigmoidal. 

Based on this behavior, we take the critical radius, $\Rc$, to be the x-value on the fractal dimension curve with the most negative slope
\begin{equation}\label{critical_radius}
    \Rc = \mathrm{argmin}_{\Rg} \frac{d \DF}{d \log_{10} (k \Rg / D)}.
\end{equation}
We evaluate \eqref{critical_radius} using simulation data by first calculating fractal dimension and its derivative with finite difference. Specifically, we use a second-order central \mynote{marg:finite_difference}{difference} at internal points and second-order forward and backward differences at the endpoints. We then fit the resulting data with cubic splines before taking the $\mathrm{argmin}$.

\subsection{Calculating the Critical Radius Scaling Exponent $\gamma$}

We now describe how we calculate the critical radius scaling exponent $\gamma$ from simulation data and contrast our approach with the one taken by previous authors. We compute $\gamma$ by finding the slope of a log-log plot of the critical radius versus the \textda number in the limit as the latter goes to zero. While conceptually straightforward, this method requires evaluating the critical radius at very small values of $\Da$. This procedure is computationally challenging for two reasons. First, as $\Da$ gets smaller, more particles are needed to observe the fractal transition, especially in 3D. $\Nc$, the number of particles in a $d$ dimensional critical cluster, scales as
\begin{equation}\label{critical_N}
    \Nc \sim (\Rc / a)^d \sim \Da^{d(\gamma - 1)},
\end{equation}
and based on analyses so far, it is clear $\gamma < 1$ \cite{meakinModelsColloidalAggregation1988, halseyElectrodepositionDiffusionLimited1990}. Second, since $\Da$ is proportional to the sticking probability in \eqref{sticking_probability}, the smaller its value, the longer each particle takes to deposit in terms of computational steps. 

As described in Sec.~\ref{si:brownian}, we used a parallel Brownian dynamics algorithm to help partially alleviate these two problems. However, calculating $\gamma$ still required substantial computational effort because the goal was always to probe deeper into the continuum limit, $\Da \to 0$. In the end, we used large-scale simulations to examine $\Da$ values where the critical clusters in 2D and 3D contained several hundred million particles, orders of magnitude more than the largest reactive deposition clusters generated previously \cite{meakinModelsColloidalAggregation1988, halseyElectrodepositionDiffusionLimited1990}.

Authors of prior studies calculated $\gamma$ using an alternative strategy. Rather than computing the critical radius directly, they instead determined $\gamma$ by exploiting data collapse \footnotetext[6]{Meakin in Ref.~\onlinecite{meakinModelsColloidalAggregation1988} actually estimated the equivalent lattice-based scaling exponent $\nu$ instead of $\gamma$, see Sec.~\ref{si:lattice}. However, the same data collapse method was used for this $\nu$ calculation.}\cite{meakinModelsColloidalAggregation1988, halseyElectrodepositionDiffusionLimited1990,Note6}. As an example of this approach, consider the relationship between the fractal dimension and the cluster radius in nondimensional form
\begin{equation}
    \DF = f( \frac{k \Rg}{D}, \Da).
\end{equation}
Taking the continuum limit and using the fact that the cluster starts off compact, $\DF = d$, yields
\begin{equation}\label{collapse}
    \DF = g(\frac{k \Rg}{\Da^{\gamma} D})
\end{equation}
where the factor $\Da^{\gamma}$ is associated with the critical radius \cite{barenblattScalingSelfsimilarityIntermediate1996}. As a result, it is possible to evaluate $\gamma$ by probing the collapse of the fractal dimension data in the low $\Da$ limit.

Such collapse-based methods, however, have several limitations. To begin with, quantitatively characterizing the extent of a collapse is not straightforward. Previous authors instead estimated $\gamma$ by judging each potential collapse by eye \cite{meakinModelsColloidalAggregation1988, halseyElectrodepositionDiffusionLimited1990}. In addition, equations such as \eqref{collapse} only apply if the \textda number is small enough. Consequently, it is necessary to leave out the highest $\Da$ value, then the next highest $\Da$ value, and so on, to assess convergence. But previous authors did not investigate convergence in a systematic manner \cite{meakinModelsColloidalAggregation1988, halseyElectrodepositionDiffusionLimited1990}. Finally, checking for a collapse is susceptible to bias if only partial data is available. For example, at certain values of $\Da$, we often cannot generate large enough clusters to see the full fractal plateau. Running a collapse calculation on such data would artificially add weight to the initial compact piece of the curve that we can successfully compute. 

In contrast, directly measuring the critical radius provides a simple way to calculate $\gamma$ and evaluate convergence. All that is necessary is to examine a log-log plot of $\Rc$ versus $\Da$. Further, direct measurement also avoids introducing any bias due to partial data.

\section{Results and Discussion}\label{sec:results}

In this section, we present the results of our two and three-dimensional Brownian dynamics simulations and evaluate whether these results are consistent with the $\gamma = 0$ prediction of the continuum model. We also briefly explore an intriguing new aspect of the CTD transition uncovered by our analysis of the critical radius. Specifically, this transition becomes increasingly sharp in the limit as $\Da \to 0$, a behavior that resembles the finite-size scaling of an equilibrium phase transition.

Simulation results for two-dimensional deposition are shown in Fig.~\ref{fig:scaling}. Panel (a) depicts the characteristic sigmoidal shape of the fractal dimension as a function of the cluster radius. At \mynote{marg:larger_values}{larger} values of the \textda number, the fractal transition occurs when the cluster has very few particles. As a result, the plateau at $\DF = 2$ due to compact growth is not yet fully visible. This lack of a compact plateau makes calculating the critical radius impossible at $\log_{10} \Da$ values larger than $-1.26$. Panel (a) also provides further evidence that the fractal dimension of the 2D dendritic morphology is approximately $1.71$ independent of the value of $\Da$ \cite{meakinModelsColloidalAggregation1988,halseyElectrodepositionDiffusionLimited1990}.

From Fig.~\ref{fig:scaling}(a), we can calculate the critical radius, \eqref{critical_radius} and evaluate its behavior in the continuum limit. Fig.~\ref{fig:scaling}(b) tracks the magnitude of the derivative of the fractal dimension. The peaks on this panel define the critical radius $\Rc$ for each $\Da$. Plotting these $\Rc$ values directly in Fig.~\ref{fig:scaling}(c) shows that the critical radius increases as the \textda number gets smaller. To find the exponent $\gamma$, in panel (d) we compute the derivative $\Gamma = \frac{d \log_{10} [k \Rc / D]}{d \log_{10} \Da}$, which converges to $\gamma$ in the limit $\Da \to 0$. This derivative gets closer to zero as $\Da$ gets smaller. Further, taking the leftmost point on the curve (dark blue) to approximate $\gamma$ yields an estimate of $-0.06$ with a $95\%$CI of $[-0.09, -0.03]$.

The confidence interval for $\gamma$ must be interpreted carefully in relation to the $\gamma = 0$ hypothesis. In particular, our calculated interval does not include zero. The remaining discrepancy can be explained by the magnitude of the \textda number. That is, for smaller values of $\Da$, $\Gamma$ might follow the trend in Fig.~\ref{fig:scaling}(d) and move closer to zero. But this trend is not robust. The bottom of the confidence interval for the dark blue point, $-0.09$, may instead indicate that the final part of the upwards drift in the data is a numerical artifact covering an underlying plateau at around $-0.1$. Such a plateau would be consistent with the position of the preceding violet, brown, and light blue points. 

We might be able to resolve this ambiguity by expanding Fig.~\ref{fig:scaling}(d) to include the next point on the left. However, to add this point, we would have to generate clusters with billions of particles at an even smaller value of $\Da$, where each particle takes more computational steps to deposit. Unfortunately, these calculations would take years of simulation time, and so are computationally prohibitive.

In spite of the concern about the trend in $\Gamma$, we can still draw two conclusions from the results of the 2D simulations. First, although previous studies reported that $\gamma \approx -0.2$, this value does not fit our data \cite{meakinModelsColloidalAggregation1988, halseyElectrodepositionDiffusionLimited1990}. Based on panel (d), it is clear that $\gamma \geq -0.1$, a significant upward revision. Second, while it is not possible to distinguish between a plateau at around $\gamma = -0.1$ and a continued trend towards $\gamma = 0$, the latter is still entirely consistent with the simulation results. Consequently, the data provides the first substantive numerical evidence that the continuum model reproduces the continuum limit of the particle-based deposition process.

\begin{figure*}
    \includegraphics{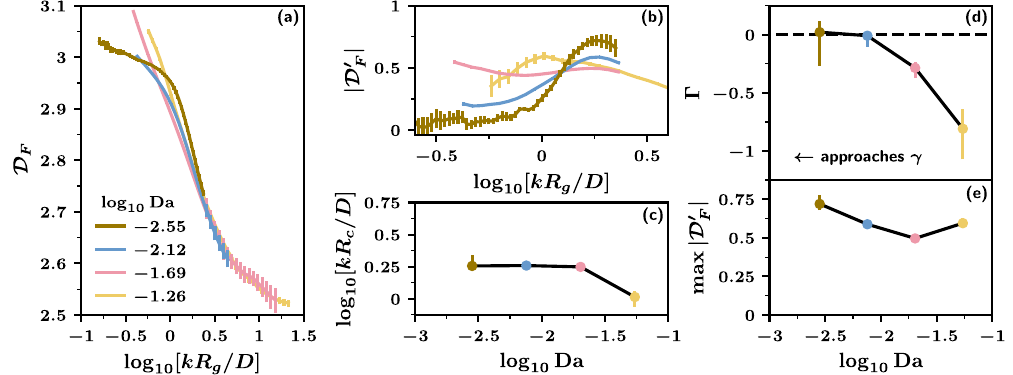}
    \caption{\label{fig:3d_scaling} Continuum limit scaling of the compact-to-dendritic transition in three dimensions. Panel axis labels are the same as in Fig.~\ref{fig:scaling}, but note the difference in scale between panel (d) and Fig.~\ref{fig:scaling}(d). The repeated caption is provided for convenience. See Table \ref{tab:3d_scaling} for further trajectory details. (a) Fractal dimension $\DF$ as a function of the radius of gyration $\Rg$ for different \textda numbers, $\Da$. (b) The absolute value of the derivative of panel (a), $|\DF'|$. (c) The critical radius $\Rc$ versus $\Da$. (d) Derivative of the log of the critical radius with respect to the log of the \textda number, $\Gamma$. The black dashed line corresponds to the linear stability analysis prediction that $\gamma = 0$. (e) Height of the peak in panel (b), $\max|\DF'|$, versus the \textda number.}
\end{figure*}

We now examine the continuum limit of the CTD transition for three-dimensional deposition. The results of our 3D simulations are shown in Fig.~\ref{fig:3d_scaling} using the same format as the 2D results. Before turning our focus to $\gamma$, we note that panel (a) is consistent with having $\DF \approx 2.5$ independent of the value of the \textda number. The critical radius is plotted in panel (c). When $\Da$ tends towards zero, this radius exhibits an immediate plateau at around $3 D / k$. Again using the leftmost (brown) point of the derivative in panel (d), we find that $\gamma = 0.02$ with a $95\%$CI of $[-0.27, 0.12]$. The confidence interval is broad due to the limited number of trajectories generated at $\log_{10} \Da = -2.55$. To increase the precision, we can instead estimate $\gamma$ using the second-to-last point from the left (light blue), which has a value of $-0.01$ with a $95\%$CI of $[-0.11, 0.02]$. However, since we are using a larger \textda number, this calculated $\gamma$ value may not be as representative of the continuum limit.

Because simulations in 3D are more expensive than in 2D, analyzing the $\gamma = 0$ hypothesis using the 3D data presents its own set of challenges. To begin with, in 3D, we could not observe the CTD transition at as wide a range of \textda numbers, and as a result, the convergence to the $\Da \to 0$ limit is not as robust. For example, unlike in 2D, the plateau that corresponds to the compact growth regime $\DF = 3$ is not fully developed in Fig.~\ref{fig:3d_scaling}(a) even for the smallest value of $\log_{10} \Da = -2.55$ (brown curve). In addition, since we generated fewer trajectories overall, the error bars in Fig.~\ref{fig:3d_scaling} are wider than in Fig.~\ref{fig:scaling}. For this reason, although the flattening trend in the 3D data is clear, it is not possible to rule out a plateau at a $\gamma$ value between zero and $-0.1$.

Despite these limitations, the estimate for $\gamma$ in 3D is consistent with the 2D estimate, $95\%$CI $[-0.09, -0.03]$. Further, taken together, the 2D and 3D data sets offer significant evidence that $\gamma = 0$ as predicted by the continuum linear stability analysis. Both $\gamma$ confidence intervals are close to zero, and the remaining differences can be plausibly explained by statistical and the \textda number convergence effects.

Our findings clarify the continuum limit of the Brownian particle deposition process in three ways. First, they suggest that this limit is well-defined. Once the particles become small enough, their exact size ceases to affect the system, and the critical radius instead scales with $D / k$. Second, our simulations demonstrate that the system convergences to the continuum limit only slowly. Although we reached $\Da \approx 10^{-4}$ in 2D and $\Da \approx 10^{-3}$ in 3D, we still did not yet observe perfect asymptotic scaling. Finally, our calculated values of $\gamma$ imply that the continuum limit of the particle-based process is accurately represented by the continuum model. That is, this model, though approximate, nevertheless preserves the dendrite formation physics.

To conclude this section, we briefly examine a novel feature of the Brownian particle system that was revealed as a side effect of our analysis of the critical radius. Specifically, the height of the peak in the fractal dimension derivative, $|\DF'|$, diverges as $\Da \to 0$ in Figs.~\ref{fig:scaling}(b) and \ref{fig:3d_scaling}(b). This behavior is notable because as $\Da$ gets smaller, the number of particles in the critical cluster, $\Nc$ in \eqref{critical_N}, goes to infinity. The continuum limit of the CTD transition is thus reminiscent of the finite-size scaling of an equilibrium phase transition \cite{newmanMonteCarloMethods1999}.

Motivated by this connection, we characterize the functional form and prefactor of the CTD transition's divergence in 2D as if we were following the standard equilibrium protocol \cite{newmanMonteCarloMethods1999}. Fig.~\ref{fig:scaling}(e) shows the height of the peak in the derivative, $\max|\DF'|$, as a function of the \textda number. The line of best fit suggests that this quantity diverges logarithmically with a prefactor of $-0.15$ ($95\%$CI of $[-0.16, -0.14]$). However, since the range of values in the figure is relatively small, it is not possible to distinguish between a logarithmic and a power-law divergence. The latter case leads to a best-fit exponent with the same value as the logarithmic prefactor, $-0.15$, with a $95\%$CI of $[-0.16, -0.14]$. We also plot the maximum value of the derivative, $\max|\DF'|$, versus $\Da$ for 3D deposition in Fig.~\ref{fig:3d_scaling}(e). Though we need more points to evaluate the prefactor, the height of the peak appears to increase in a comparable fashion in this dimension.

Beyond a numerical characterization of the divergence of $|\DF'|$, many aspects of the finite-size scaling of the CTD transition still need to be clarified. For example, how does this divergence emerge from the microscopic dynamics? And how exactly does the behavior of the transition relate to finite-size scaling in equilibrium? 

\section{Conclusion}\label{sec:conclusion}

In this work, we examined the continuum (small \textda number) limit of the dendritic deposition of Brownian particles. Specifically, we compared Brownian dynamics simulations of the compact-to-dendritic transition in the limit as $\Da \to 0$ against the predictions of a continuum linear stability analysis. From the simulations, we estimated that the critical radius scaling exponent $\gamma$ is $-0.06$ with a $95\%$CI of $[-0.09, -0.03]$ in 2D and $ -0.01$ with a $95\%$CI of $[-0.11, 0.02]$ in 3D. These values do not exactly reproduce the $\gamma = 0$ scaling derived from the stability analysis. However, the remaining discrepancy can be explained by convergence effects. Our results thus show, in contrast to previous studies, that the continuum limit of the particle-based system is consistent with the continuum deposition model.

Based on these findings, we expect that continuum models will also faithfully capture the dynamics of more complex deposition processes. Such models are already widely used to study electrodeposition and lithium metal batteries \cite{aogakiImageAnalysisMorphological1982,chazalvielElectrochemicalAspectsGeneration1990,sundstromMorphologicalInstabilityElectrodeposition1995, sanderDiffusionlimitedAggregationKinetic2000,monroeEffectInterfacialDeformation2004, monroeImpactElasticDeformation2005, tikekarStabilityAnalysisElectrodeposition2014, liuStabilizingLithiumMetal2016,tuNanoporousHybridElectrolytes2017, choudhuryConfiningElectrodepositionMetals2018}. Consequently, our results provide a retrospective theoretical validation of this work.

There is, however, one important caveat regarding the accuracy of the continuum approximation. Despite reaching \textda numbers on the order of $10^{-4}$ in 2D and $10^{-3}$ in 3D, we did not observe perfect $\gamma = 0$ convergence, meaning that the Brownian particle system approaches its continuum limit quite slowly. This convergence behavior will likely extend to other deposition processes. Accordingly, when using a continuum model of dendritic deposition, it is essential to be aware of the value of the \textda number or the equivalent nondimensional measure of the particle size. Unless this parameter is very small, the continuum approximation itself may be a significant source of error.

\section{Acknowledgments}

We thank Steve Whitelam, Tomislav Begu\v si\'c, and especially Emiliano Deustua for providing comments on the manuscript. DJ acknowledges support from the Department of Energy Computational Science Graduate Fellowship under Contract No. DE-FG02–97ER25308. This work was supported by a grant from NIGMS, National Institutes of Health, (R01GM125063) to TFM. 

\bibliography{references}

\end{document}

\preprint{APS/123-QED}

\title{Supplemental Material: Continuum Limit of Dendritic Deposition}

\author{Daniel Jacobson}
\email{jacobson.daniel.r@gmail.com}
\affiliation{Division of Chemistry and Chemical Engineering, California Institute of Technology, Pasadena, California 91125, USA}
\author{Thomas F. Miller III}
\email{tom@iambic.ai}
\altaffiliation[Now at ]{Iambic Therapeutics, San Diego, CA, USA}
\affiliation{Division of Chemistry and Chemical Engineering, California Institute of Technology, Pasadena, California 91125, USA}

\date{\today}

\maketitle

\section{Linear Stability Analysis of the Continuum Deposition Model}\label{si:linear}

The continuum reactive deposition model was first characterized by Witten and Sander in Ref.~\onlinecite{wittenDiffusionlimitedAggregation1983}. In this section, we reproduce their analysis using the updated conceptual framework from the main text \footnotetext[0]{Witten and Sander's results were originally expressed in terms of the equivalent parameters from their lattice-based deposition simulations, see Sec.~\ref{si:lattice} and Ref.~\onlinecite{erbanReactiveBoundaryConditions2007}.}\cite{Note0}. We focus, in particular, on two-dimensional (2D) deposition. The analysis for three-dimensional deposition is similar.

As introduced in Sec.~\ref{sec:model}, the concentration field $C$ that surrounds the cluster in the continuum model is determined by the diffusion equation and boundary conditions
\begin{equation}\label{continuum}
\begin{gathered}
    \nabla^2 C = 0, \\
    C(|\pos| = b) = C_0, \\
  	D \nabla C(\posB) \cdot \normal(\posB) = k C(\posB).
\end{gathered}
\end{equation}
Here, $\pos$ is position, $b$ is the bath radius, $C_0$ is the bath concentration, $\posB$ is a position on the cluster boundary, $D$ is the diffusion constant, and $k$ is the surface reaction rate constant. The value of $C_0$ does not affect the analysis of the growth behavior as long as $C$ is pseudosteady (see Sec.~\ref{sec:model} of the main text). We want to understand the growth behavior in the limit as $b \to \infty$. While it is possible to analyze the model as written and then take this limit, it is simpler to take the limit first. The generalization of the bath boundary condition to an infinite domain in 2D is \cite{sanderFractalsFractalCorrelations1994} 
\begin{equation}
    C \to C_1 \ln(r / r_0) + C_2 \text{ as } r \to \infty.
\end{equation}
In this equation, $r$ is the radial polar coordinate, $C_1$ is a known constant that plays the same role as $C_0$ in \eqref{continuum}, and the parameters $r_{0}$ and $C_{2}$ are linked undetermined constants. 
    
Following Witten and Sander, the growth behavior of the model can be characterized using a linear stability analysis \cite{wittenDiffusionlimitedAggregation1983}. This analysis takes the cluster boundary to be composed of a circle plus a small cosine perturbation with a given wavelength. If the calculated growth rate of the perturbation is faster than the growth rate of the cluster radius, then deposition will be unstable. Analyzing a cosine perturbation is enough to understand the stability of any small boundary perturbation for two reasons. First, any perturbation can be decomposed into a Fourier series, and second, the stability analysis equations are linear.

To begin, we present the equations that describe the perturbed cluster. In polar coordinates, the boundary of this cluster $\zeta(\theta)$ is 
\begin{equation}\label{boundary}
    \zeta(\theta) = \zeta_{0} + \epsilon_{m} \cos(m \theta).
\end{equation}
Here, $\zeta_{0}$ is the cluster radius, $\epsilon_{m}$ is the size of the perturbation and $m \in \{1, 2, 3, ...\}$. We then nondimensionalize with 
\begin{equation}
\begin{split}
  \bar{\zeta} &= \zeta / \zeta_{0}, \\
  \bar{\epsilon}_m &= \epsilon_m / \zeta_{0}, \\
  \bar{\nabla} &= \zeta_{0} \nabla,
\end{split}
\qquad
\begin{split}
  \bar{C} &= C / C_{1}, \\
  \bar{r} &= r / \zeta_{0}, \\
  \psi &= \frac{k \zeta_{0}}{D}.
\end{split}
\end{equation}
As a result, \eqref{boundary} becomes
\begin{equation}
  \bar{\zeta}(\theta) = 1 + \bar{\epsilon}_{m} \cos(m \theta),
\end{equation}
and the diffusion equation and boundary conditions, \eqref{continuum}, become
\begin{equation}\label{nondim_continuum}
\begin{gathered}
  \bar{\nabla}^2 \bar{C} = 0, \\
   \bar{C} \to \ln(\bar{r}) + \bar{C}_3 \text{ as } \bar{r} \to \infty, \\
  \bar{\nabla} \bar{C}(\bar{\zeta}(\theta), \theta) \cdot \boldsymbol{n} = \psi \bar{C}(\bar{\zeta}(\theta), \theta),
\end{gathered}
\end{equation}
where $\bar{C}_3$ is a new undetermined constant resulting from the transformation. 

Next, we solve for $\bar{C}$ perturbatively up to first order in the small parameter $\bar{\epsilon}_{m}$. Specifically, we seek a solution of the form \cite{dykePerturbationMethodsFluid1975}
\begin{equation}
	\bar{C}(\bar{r}, \theta) = \bar{C}_{I} (\bar{r}, \theta) + \bar{\epsilon}_{m} \bar{C}_{II}(\bar{r}, \theta).
\end{equation}
To find $\bar{C}_{I}$, we set $\bar{\epsilon}_{m} = 0$. Since there is no longer any perturbation, $\bar{C}_{I}$ is a function of $\bar{r}$ only, leading to the solution
\begin{equation}
 	\bar{C}_{I}(\bar{r}) = \ln(\bar{r}) + \psi^{-1}.
\end{equation}

Solving for the second term, $\bar{C}_{II}$ requires first expanding the reactive boundary condition at the surface to $O(\bar{\epsilon}_m)$. The outward surface unit normal is
\begin{equation}
\normal =  \boldsymbol{\hat{e}}_r + \bar{\epsilon}_{m} m \sin(m \theta)\boldsymbol{\hat{e}}_{\theta},
\end{equation}
where $\boldsymbol{\hat{e}}_r$ and $\boldsymbol{\hat{e}}_{\theta}$ are the $r$ and $\theta$ unit vectors, respectively. Consequently, the reactive boundary condition in \eqref{nondim_continuum} reduces to
\begin{equation}
\partial_{\bar{r}} \bar{C}(\bar{\zeta}(\theta), \theta) = \psi \bar{C}(\bar{\zeta}(\theta), \theta).
\end{equation}
After expanding each side of this equation at the boundary and gathering $O(\bar{\epsilon}_m)$ terms, we have for the $\bar{C}_{II}$ problem
\begin{equation}
\begin{gathered}
  \bar{\nabla}^2 \bar{C}_{II} = 0, \\
  \bar{C}_{II} \to 0 \text{ as } \bar{r} \to \infty, \\
  \partial_{\bar{r}} \bar{C}_{II}(1, \theta) = \psi \bar{C}_{II}(1, \theta) + \cos(m \theta)(\psi + 1),
\end{gathered}
\end{equation}
leading to the solution
\begin{equation}
  \bar{C}_{II}(\bar{r}, \theta) = - \frac{\psi + 1}{\psi + m} \frac{\cos(m \theta)}{\bar{r}^{m}}.
\end{equation}

We can now evaluate the stability of the growth. First, the (negative) flux at the surface is  
\begin{equation}
\bar{\nabla} \bar{C}(\bar{\zeta}(\theta), \theta) \cdot \normal = 1 +\frac{m - 1}{1 + m\psi^{-1}} \bar{\epsilon}_{m} \cos(m \theta),
\end{equation}
so the dimensional growth velocity, from \eqref{velocity} in the main text, is
\begin{equation}
   v(\theta) =  \frac{\mu D C_{1}}{\zeta_0^2} \left[ \zeta_0 +\frac{(m - 1) \epsilon_{m} \cos(m \theta)}{1 + m \frac{D}{k \zeta_{0}}}  \right].
\end{equation}
Comparing this expression with the equation for the boundary \eqref{boundary}, we see that the first term expresses the growth velocity of the radius $v_{\zeta_0}$ while the second term expresses the growth velocity of the perturbation $v_{\epsilon_{m}}$. As a result, the stability parameter $\chi$ is \cite{mullinsMorphologicalStabilityParticle1963}
\begin{equation}\label{stability}
\chi = \frac{v_{\epsilon_m} / \epsilon_m}{v_{\zeta_0} / \zeta_0} = \frac{m - 1}{1 + m \frac{D}{k \zeta_0}}
\end{equation}

Eq.~\eqref{stability} shows that he deposition is unstable when $\zeta_0 \gg D / k$. That is, perturbations for $m > 2$ grow faster than the radius itself. However, the instability is suppressed when $\zeta_0 \ll D/k$. Based on this analysis, Witten and Sander concluded that the critical radius for the compact-to-dendritic transition in the continuum model scales as $D/k$ \cite{wittenDiffusionlimitedAggregation1983}.

\section{Brownian Dynamics Simulations}\label{si:brownian}

\subsection{Brownian Dynamics Algorithm}

The basis for our Brownian dynamics algorithm is \eqref{gaussian_update}, the standard Gaussian update of the incoming particle's position \cite{allenComputerSimulationLiquids2017}. However, since the particle is often far from the cluster, using this kind of update exclusively is inefficient. Instead, we adopt several standard computational tricks to speed up the simulation of the dynamics \cite{sanderFractalsFractalCorrelations1994, tolmanOfflatticeHypercubiclatticeModels1989}. These tricks involve breaking the incoming particle's motion into three different types of steps: small steps, large steps, and first-hit steps. In this subsection, we describe each kind of step individually and then show how we combine them to form the complete algorithm. For a summary, see Alg.~\ref{alg:bd_combined}.

\subsubsection{Small Steps}

During a small step, we displace the incoming particle following the standard Gaussian procedure and check for contacts with particles in the supercluster. We run the contacts check by treating the displacement as straight-line motion between the initial and final positions \cite{erbanReactiveBoundaryConditions2007, singerPartiallyReflectedDiffusion2008}. Then to find the supercluster particles that intersect this path, we use the line-sphere intersection algorithm \cite{eberly3DGameEngine2006}. We also constrain the maximum distance that the particle can move in each dimension during a single step to be $\lcut$, drawing displacements from a truncated rather than a standard Gaussian distribution. Bounding the displacement limits the number of possible contacts per step and allows us to quickly check these contacts using a cell list \cite{allenComputerSimulationLiquids2017}. Further, the bias introduced by this constraint is negligible as long as $\lcut$ is large.

When the incoming particle does make contact with the supercluster, we draw a random number and compare it to the sticking probability, \eqref{sticking_probability}, as described in the main text. If the particle does not stick, it bounces off the supercluster such that the incoming and outgoing angles of reflection are equal \cite{erbanReactiveBoundaryConditions2007, singerPartiallyReflectedDiffusion2008}. After this contact, the particle continues to travel in a straight line, bouncing off the supercluster again if necessary. The step ends when the incoming particle either sticks or the lengths of its straight-line displacements between reflections sum to the magnitude of the initial displacement. Note that in the future, we plan to implement the additional sticking probability correction described in Ref.~\onlinecite{boccardoImprovedSchemeRobin2018}. This correction allows a higher level of accuracy to be achieved for the same fixed value of the timestep.

Small steps have issues with numerical stability in two rare situations. The first situation occurs when the final position of the incoming particle is very close to the supercluster boundary. Such a configuration can cause the contact-checking algorithm to miss the nearby boundary during the next step. We prevent this problem by checking if the incoming particle ends up within $2a + \iota$ of any supercluster particles. Here $\iota / a = 1 \times 10^{-8}$ is a threshold parameter that we use throughout the algorithm. If the incoming particle is within this distance, we treat its final position as a contact and make a draw against the sticking probability. We then either attach the particle to the cluster or, when the draw fails, reject the original move such that the particle position $\posP(t+\dt) = \posP(t)$. 

The second situation that leads to numerical instability is similar to the first. It arises when the incoming particle bounces off the supercluster at a cusp within $\iota$ of the boundary of another supercluster particle. In this case, the contact-checking algorithm may once again miss the nearby boundary during the next cycle of motion. We avoid this problem by rejecting the move rather than continuing to resolve the bounce. 

\subsubsection{Large Steps}\label{si:large_steps}

We use the second type of step, large steps, when the incoming particle is far from the supercluster. During these steps, we first find the distance $H$ from the particle to the nearest supercluster particle. This operation can be done in $O(\ln N)$ time using a $k$-d tree \cite{tolmanOfflatticeHypercubiclatticeModels1989, blanco2014nanoflann}. We then displace the incoming particle by $H - 2a - 2 \iota$ in a random direction. Such a step size guarantees that no contacts will occur. Note that we must subtract $2 \iota$ instead of just $\iota$ from displacement since supercluster particles already have a skin of size $\iota$ for collision detection, as mentioned previously.

\subsubsection{First-Hit Steps}\label{si:first-hit_steps}

The final type of step we use in the Brownian dynamics algorithm is the first-hit step. For this step, we first define a circle (or sphere in 3D) centered at the origin that bounds the supercluster. Then, if the incoming particle ends up outside this region, we return it to a point on the bounding surface by integrating the dynamics analytically \cite{sanderFractalsFractalCorrelations1994}. The analytical integration involves calculating the first-hit distribution, the probability distribution that the particle hits the bounding circle (or sphere) for the first time at a given point. The 2D first-hit distribution is derived in Ref.~\onlinecite{sanderFractalsFractalCorrelations1994}, and we derive the 3D first-hit distribution in Sec.~\ref{si:first-hit}. 

A key feature that makes the first-hit calculation possible is that in the limit as the bath distance goes to infinity, the effect of the bath can be ignored. In other words, the analytical first-hit distribution can be calculated in an unbounded domain. To understand how the unbounded domain approximation works, consider the deposition dynamics when the bath is in place. In this case, if the incoming particle makes it all the way out to the bath, it is killed. The next diffusive particle that is launched is then equally likely to approach the supercluster from any direction. Now consider the case without a bath. When the incoming particle reaches the bath distance, it will continue to diffuse rather than be killed. However, at this point, the particle is so far from the supercluster that when it returns, it will also be equally likely to approach from any direction. Consequently, the system with a bath and the system without a bath have the same first-hit distribution.

\subsubsection{Combined Algorithm}\label{si:combine_steps}

Having specified the three types of steps, we combine them to form the Brownian dynamics algorithm according to Alg.~\ref{alg:bd_combined}. First, we check the cell list to see if any supercluster particles are in the incoming particle's current cell or any adjacent cells. If so, we choose to take a small step. Otherwise, we take a large step. Following this choice, if the incoming particle ends up outside the circle or sphere that bounds the supercluster, we take a first-hit step. Lastly, when launching a new particle, we initialize it from a random point on the same first-hit bounding surface.

\begin{algorithm}[H]
    \caption{Brownian Dynamics Algorithm}
    \label{alg:bd_combined}
    \begin{algorithmic}
    \While{cluster size $<$ desired size}
        \State initialize new incoming particle
        \While{incoming particle not stuck}
            \If{current and adjacent cells not empty}
                \State take small step
            \Else
                \State take large step
            \EndIf
            \If{outside circle/sphere that bounds cluster}
                \State take first-hit step
            \EndIf
        \EndWhile
    \EndWhile
    \end{algorithmic}
\end{algorithm}

\subsection{Simulation Convergence}\label{si:convergence}

Since large steps and first-hit steps are exact, the convergence of the Brownian dynamics algorithm is controlled by the dynamics of the small (Gaussian) steps. These steps have two convergence parameters, the timestep $\dt$ and the cutoff distance $\lcut$, and they reproduce the target dynamics in the limit as $\dt \to 0$ and $\lcut \to \infty$ \cite{erbanReactiveBoundaryConditions2007, singerPartiallyReflectedDiffusion2008}. Consequently, running simulations with a nonzero timestep and a finite cutoff distance introduces a systematic bias in the calculation of the critical radius. In this section, we discuss how we selected values of $\dt$ and $\lcut$ that are converged. That is, where the systematic bias associated with the parameter choice is substantially smaller than the statistical error in the critical radius. We also examine how convergence relates to the computational efficiency of the simulation.

We begin with this latter topic by considering how the timestep and the cutoff distance affect the wall-clock speed of our calculations. The starting point for this analysis is the version of the position update equation \eqref{gaussian_update} that we use in our simulations that implements the cutoff
\begin{equation}\label{truncated_gaussian_update}
    \posP(t + \dt) = \posP(t) + \eta \boldsymbol{g}_{T}(\barlcut).
\end{equation}
Here, $\boldsymbol{g}_{T}(\barlcut)$ is a truncated Gaussian distribution that is zero after $\barlcut$ standard deviations and $\eta = \sqrt{2 D \dt}$ is a prefactor that controls the step size. Eq.~\eqref{truncated_gaussian_update} is conceptually convenient because the shape of the particle displacement distribution depends only on $\barlcut$.

Given \eqref{truncated_gaussian_update}, how does changing $\barlcut$ and $\eta$ (which we work with in place of $\dt$) affect the simulation speed? First, consider making $\barlcut$ larger at a fixed value of $\eta$. This change increases the dimensional cutoff $\lcut = \barlcut \eta$, the maximum possible jump size of the incoming particle. Consequently, it also increases the number of cluster particles that the incoming particle might encounter on each step. Sorting through these additional potential collisions slows down the simulation. 

More surprising behavior results from making $\eta$ larger for a fixed value of $\barlcut$. This change increases the sticking probability \eqref{sticking_probability} and the distance the particle jumps per step, making the simulation faster. However, it also increases the dimensional cutoff $\lcut = \barlcut \eta$, slowing down the simulation due to the greater number of potential collisions per step that must be analyzed. In particular, the relative magnitudes of these effects depend on the system parameters. When $\eta$ is small, the impact of increasing the sticking probability and the displacement per step dominates. Consequently, increasing $\eta$ makes the simulation faster. In contrast, when $\eta$ is already large, the impact of adding more potential collisions dominates. As a result, increasing $\eta$ makes the simulation \textit{slower}. It follows that the simulation speed at fixed $\barlcut$ is a concave function of $\eta$ with a global maximum value $\eta^*$. Empirical tests show that $\eta^*$ gets larger as $\barlcut$ gets smaller. But in general, the location of $\eta^*$ may also depend on the \textda number, the implementation of the simulation code, and the architecture of the computer cluster.

\begin{figure}
  \includegraphics{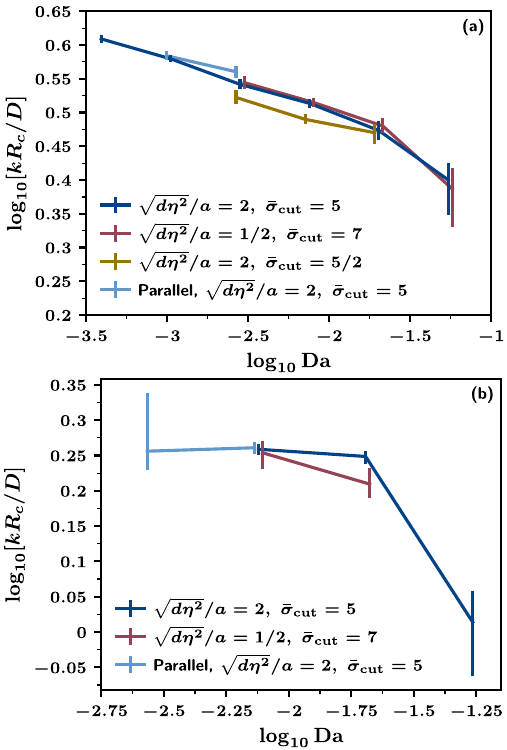}
  \caption{\label{fig:combined_convergence} Test of the timestep, cutoff, and parallel algorithm convergence for the 2D (a) and 3D (b) Brownian dynamics simulations. Each panel shows the critical radius $\Rc$ versus the \textda number $\Da$ for the serial (dark blue, brown, and violet curves) and parallel (light blue curve) algorithms. The serial curves were generated with different values of the cutoff $\barlcut$ (see \eqref{truncated_gaussian_update}) and the timestep $\sqrt{d\eta^2} = \sqrt{2dD\dt}$, where $d$ is the dimension. Note that the x-values of the violet, brown, and light blue curves are slightly shifted to increase the visibility of the individual points.}
\end{figure}

Having clarified the effect of $\eta$ and $\barlcut$ on the simulation wall-clock time, we now describe how we searched for efficient, converged parameters. Calculating the critical radius is computationally expensive. For this reason, we sought to evaluate convergence at high $\Da$ before running our low $\Da$ calculations. Even then, we were limited to sampling only a small part of the grid of possible pairs of $\eta$ and $\barlcut$ values. In 2D, we started with an initial guess of $\sqrt{d \eta^2} / a = 2, \ \barlcut = 5$ ($\sqrt{d \eta^2}$ is the root mean square displacement per timestep in $d$ dimensions when $\barlcut = \infty$). The critical radius values for this choice of parameters are plotted in Fig.~\ref{fig:combined_convergence}(a) as the dark blue curve. To test for convergence, we compared against calculations run with $\sqrt{d \eta^2} / a = 1/2, \ \barlcut = 7$, the violet curve, which took about five times as long. The close overlap of the error bars for both parameter sets suggests that our initial guess was converged. Further runs at $\sqrt{d \eta^2} / a = 2, \ \barlcut = 7$ (not shown) confirmed that this overlap is genuine and is not the result of a cancellation of errors.

We then explored whether variations on our initial guess could improve the speed of our simulation. First, we tested halving the cutoff, $\sqrt{d \eta^2} / a = 2, \ \barlcut = 5/2$, which is plotted as the brown curve in Fig.~\ref{fig:combined_convergence}(a). While at the lowest values of $\Da$, our simulations ran about $50\%$ faster than with $\barlcut = 5$, the smaller cutoff did not yield a converged critical radius. We next tried increasing the value of $\sqrt{d \eta^2} / a $ to $8$. But with this larger $\eta$, our simulations still took roughly the same amount of wall-clock time, making the convergence not worth evaluating. Since our search did not yield a more efficient set of parameters, we used our initial guess $\sqrt{d \eta^2} / a = 2, \ \barlcut = 5$ for the 2D simulations presented in the main text.

Continued exploration would have likely yielded more efficient parameter values. In particular, our final choice of $\barlcut = 5$ was probably conservative. Although the $\barlcut = 5/2$ (brown) curve is not converged, the deviation is modest, suggesting that intermediate cutoff values may perform well. In addition, while increasing the timestep did not lead to any speed gains at $\barlcut = 5$, it could be more effective for smaller cutoffs. Any potential efficiency gains, however, had to be weighed against the cost of the additional exploration calculations. 

We tested the convergence of the 3D simulations using a similar approach as in 2D. But since calculating the critical radius is even more expensive in this dimension, our exploration of parameter space was further limited. We started with the same set of parameters that worked well in 2D,  $\sqrt{d \eta^2} / a = 2, \ \barlcut = 5$. Fig.~\ref{fig:combined_convergence}(b) shows that these parameters (dark blue curve) exhibit slightly worse convergence than in Fig.~\ref{fig:combined_convergence}(a). At a larger value of $\Da$, the blue curve does not quite overlap with the  $\sqrt{d \eta^2}/a = 1/2, \barlcut = 7$ (violet) curve. Nevertheless, since we still found convergence at smaller values of $\Da$, we again used $\sqrt{d \eta^2}/a = 2, \barlcut = 5$ for the 3D simulations presented in the main text. We also confirmed that making $\eta$ smaller slowed down the simulation at all values of $\Da$. Consequently, this section of parameter space did not offer any trade-off-free increases in accuracy or speed.

\subsection{Parallel Algorithm}

Even with computational tricks like large and first-hit steps, generating large clusters is still expensive, especially in 3D. Consequently, to further increase the speed of our simulations, we used a simple parallel algorithm. In this algorithm, the dynamics of $M$ particles are propagated simultaneously on $M$ different CPUs. To begin, these particles each take a total of $s$ small or large steps (first-hit steps are also taken as needed). If a particle sticks to the cluster during this interval, the remaining steps are forfeited. After these $s$ steps, information about which particles deposited onto the cluster is shared between the processors and new incoming particles are introduced to the system. In rare cases, two incoming particles stick in the same place, or an incoming particle ends up inside a newly added supercluster particle. To resolve these issues, we eliminate one of the offending particles. If one particle is free (versus attached to the cluster), we eliminate that one. Otherwise, we choose randomly.

The parallel interval parameter $s$ offers the following trade-off. Using a small value of $s$ decreases the speed of the algorithm because it increases the amount of inter-processor communication. However, using a large value of $s$ is inefficient because an incoming particle may stick early in the cycle. In this case, the associated processor must wait a long time before it can be occupied with a new incoming particle. A large value of $s$ also causes a higher fraction of the incoming particles to stick in each cycle, creating more overlap artifacts. For our simulations, we used $s = 100$.

While the parallel algorithm speeds up the generation of large clusters, it also biases the growth morphology. The easiest way to limit this bias is to use the parallel algorithm for less time by starting from a medium-sized, serially-generated cluster rather than a single seed particle \cite{kaufmanParallelDiffusionlimitedAggregation1995}. Here, if we estimated that the compact-to-dendritic transition would occur at $\Nc$ particles, we switched to the parallel algorithm at $\Nc / 2$ particles. Reducing $M$ also decreases the algorithm's bias. However, we set $M = 16$ to take advantage of our computer cluster's architecture.

Fig.~\ref{fig:combined_convergence}(a) shows an empirical test of the parallel bias for 2D deposition. This figure compares the critical radius computed with the serial algorithm (dark blue curve) and the parallel algorithm (light blue curve). Though there are only two data points, the bias appears to be smaller for smaller values of $\Da$.

We used the same setup in Fig.~\ref{fig:combined_convergence}(b) to test the bias of the parallel algorithm in 3D. Comparing the serial and parallel algorithms at $\log_{10} \Da = -2.12$ shows that the bias introduced by the latter at this \textda number is negligible. Based on this data point and the trend of the bias with $\Da$ in 2D, we used the parallel algorithm for our 3D simulations at $\log_{10} \Da = -2.55$. We did not use the parallel algorithm for our 2D simulations. 

\begin{table}
\caption{\label{tab:scaling} Details of the 2D deposition simulations used to generate Fig.~\ref{fig:scaling} in the main text. The first column lists the \textda number $\Da$. The second column, $\Rc$ Runs, indicates the number of runs used to calculate the critical radius. A smaller number of runs listed in the next column, Extended Runs, were extended in some cases to map out the fractal dimension curve. Finally, the last column indicates whether we used the parallel or serial algorithm.}
\begin{ruledtabular}
\begin{tabular}{dlll} 
\log_{10} \Da & $\Rc$ Runs & Extended Runs & Parallel \\
\hline
-1.26 & 400,000 & 100 & No \\
-1.69 & 40,000 & 100 & No \\
-2.12 & 10,000 & 100 & No \\
-2.55 & 1000 & 100 & No \\
-2.98 & 400 & N/A & No \\
-3.14 & 400 & N/A & No \\
\end{tabular}
\end{ruledtabular}

\caption{\label{tab:3d_scaling} Descriptions of the 3D simulations used to generate Fig.~\ref{fig:3d_scaling} in the main text. The columns are the same as in Table \ref{tab:scaling}.}
\begin{ruledtabular}
\begin{tabular}{dlll} 
\log_{10} \Da & $\Rc$ Runs & Extended Runs & Parallel \\
\hline
-1.26 & 10,000 & 100 & No \\
-1.69 & 30,000 & 10 & No \\
-2.12 & 1000 & 10 & No \\
-2.55 & 10 & N/A & Yes \\
\end{tabular}
\end{ruledtabular}
\end{table}

\subsection{Trajectory Details}

Tables \ref{tab:scaling} and \ref{tab:3d_scaling} summarize the number of 2D and 3D trajectories (runs) used to generate the figures in the main text. These tables include two entries for the number of runs for each value of $\Da$, $\Rc$ runs, and extended runs. The $\Rc$ runs indicate the number of trajectories we used to calculate the critical radius. After this calculation, we continued with a smaller number of runs, extended runs, to map out the rest of the fractal dimension curve. The switch between the $\Rc$ runs and extended runs is also the reason that the size of the error bars increases substantially in Figs.~\ref{fig:scaling}(a) and \ref{fig:3d_scaling}(b) in the main text when the cluster radius of gyration exceeds the critical radius, $\Rg > \Rc$.

\section{Reactive Deposition on a Lattice}\label{si:lattice}

While the main text focuses on the dendritic deposition of Brownian particles in continuous space, several authors have also examined what happens when these particles are instead confined to a lattice \cite{wittenDiffusionlimitedAggregation1983, meakinDiffusioncontrolledClusterFormation1983,meakinDiffusioncontrolledDepositionFibers1983,meakinModelsColloidalAggregation1988}. In this section, we briefly review these efforts and discuss how they relate to our analysis of the critical radius scaling exponent $\gamma$.

When the particles diffuse on a lattice, the Brownian deposition system shows an equivalent compact-to-dendritic transition \cite{wittenDiffusionlimitedAggregation1983, meakinDiffusioncontrolledClusterFormation1983,meakinModelsColloidalAggregation1988}. In this case, the critical radius depends on the sticking probability $P$, which, unlike in continuous space, is a well-defined physical parameter. When $P$ approaches $1$, the system reduces to diffusion-limited aggregation on a lattice with an $\Rc$ that is the size of a single particle. In the opposite limit, $P \to 0$, the critical radius exhibits a nontrivial scaling \cite{meakinModelsColloidalAggregation1988, meakinFractalsScalingGrowth1998}
\begin{equation}\label{lattice_scaling}
    \Rc / l \sim P^\nu 
\end{equation}
where $l$ is the lattice constant. Meakin used simulations to estimate that, for a triangular lattice, $\nu$ is equal to $-1.2$ \cite{meakinModelsColloidalAggregation1988, meakinFractalsScalingGrowth1998}. 

What then can lattice-based dendritic deposition tell us about dendritic deposition in continuous space? One might expect that the lattice has a minimal impact on dendrite formation. After all, in surface growth models, the on- versus off-lattice distinction does not change the large-length scale emergent properties of the morphology \cite{barabasiFractalConceptsSurface1995}. It is now clear, however, that the structure of dendritic deposition is influenced by the underlying spatial geometry. Specifically, the presence of a lattice with a given symmetry introduces persistent anisotropy into the cluster's fractal form \cite{grebenkovHowAnisotropyBeats2017}. 

Despite this anisotropy, the lattice and continuous space versions of Brownian particle deposition may still share the same critical radius scaling behavior. We can develop this hypothesis further by appealing to a rigorous connection that exists between the two systems. In particular, it is possible to use a lattice model to propagate the exact deposition dynamics of the \textit{continuous space} Brownian particle deposition process. The continuous space dynamics emerge in the limit as the lattice constant $l$ goes to zero while the particle radius $a$ remains fixed \cite{erbanReactiveBoundaryConditions2007}. Fig.~\ref{si:lattice} demonstrates that in this limit, particles start to take up more and more lattice sites. However, similar to taking $\dt \to 0$ in an off-lattice simulation, when taking $l \to 0$ on a lattice, the sticking probability $P$ acts as a second convergence parameter. The rate constant $k$ being simulated depends on $l$ and $P$ through
\begin{equation}\label{k_lattice}
    k = D P / l.
\end{equation}
Therefore, to maintain a constant surface reaction rate as $l \to 0$, the ratio $P/l$ must be kept fixed. 

We can use \eqref{k_lattice} to translate lattice-based results into continuous space equivalents. Following this strategy, we find that Meakin's estimate for the critical radius scaling exponent on a triangular lattice, $\nu = -1.2$, yields $\gamma = -0.2$ \cite{meakinModelsColloidalAggregation1988, meakinFractalsScalingGrowth1998}. In the future, lattice simulations coupled with \eqref{k_lattice} may also offer a convenient approximate means to access small values of the \textda number that would be too expensive to study in proper continuous space.

\begin{figure}
  \includegraphics{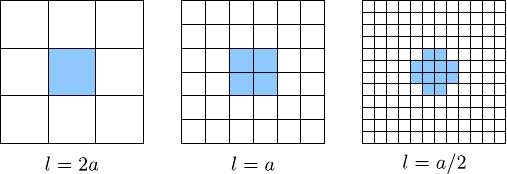}
  \caption{\label{fig:lattice} Illustration of a particle (light blue) with radius $a$ represented on square lattices with varying lattice constants $l$. As $l$ becomes smaller, the particle takes up an increasing number of lattice sites.}
\end{figure}

\section{First-Hit Distribution in Three Dimensions}\label{si:first-hit}

Here, we calculate the (3D) first-hit distribution for a Brownian particle released from a point outside an absorbing sphere. This distribution gives the probability density that the particle first makes contact with the sphere at a given position on the surface. Our derivation closely follows the derivation of the 2D first-hit distribution found in Ref.~\onlinecite{sanderFractalsFractalCorrelations1994}. We also describe how we sample the 3D first-hit distribution for our deposition simulations. 

\begin{figure}
  \includegraphics[scale=1.6]{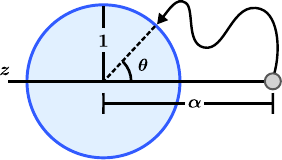}
  \caption{\label{fig:first_hitting_diagram} The particle lies on the $z$-axis a distance $\alpha$ away from the center of an absorbing unit sphere. Once released, it diffuses until it escapes to infinity or absorbs at polar angle $\theta$.}
\end{figure}

To set up the calculation, consider a sphere with a given radius centered at the origin and a Brownian point particle outside of this sphere as shown in Fig.~\ref{fig:first_hitting_diagram}. After nondimensionalizing by the sphere's radius, the particle position vector is $\bar{\pos}_{0} = \alpha \boldsymbol{\hat{e}}_z$ where $\alpha > 1$. Since the geometry is symmetric in the azimuthal angle, the first-hit distribution is a function of the polar angle $\theta$ alone.

To determine the first-hit distribution, we start with the dimensionless probability density of the particle $\xi(\bar{\pos}, \tau)$. This density evolves according to the diffusion equation
\begin{equation}\label{diffusion_equation}
  \frac{\partial \xi}{\partial \tau} =  \bar{\nabla}^{2} \xi.
\end{equation}
The boundary condition for $\xi$ on the sphere is 
\begin{equation}\label{xi_boundary}
  \xi(|\bar{\pos}| = 1, \tau) = 0
\end{equation}
and the initial condition is
\begin{equation}
  \xi(\bar{\pos}, 0) = \delta(\bar{\pos} - \bar{\pos}_{0}).
\end{equation}

Instead of solving for $\xi$ explicitly, we integrate \eqref{diffusion_equation} with respect to time, yielding
\begin{equation}
    \bar{\nabla}^{2} \int_{0}^{\infty} \xi d\tau = - \delta(\bar{\pos} - \bar{\pos}_{0}).
\end{equation}
The right-hand side of this equation follows from
\begin{equation}\label{infinite_time}
  \xi(\bar{\pos}, \tau = \infty) = 0,
\end{equation}
which is justified since the particle probability density eventually either absorbs or spreads across the infinite domain making $\xi$ arbitrarily small. After defining $\Xi(\bar{\pos}) = \int_{0}^{\infty} \xi(\bar{\pos}, \tau) d\tau$, we have 
\begin{equation}\label{electrostatics}
    \bar{\nabla}^{2} \Xi = - \delta(\bar{\pos} - \bar{\pos}_{0})
\end{equation}
with the boundary condition 
\begin{equation}\label{Xi_boundary}
  \Xi(|\bar{\pos}| = 1) = 0.
\end{equation}

Eqs.~\eqref{electrostatics} and \eqref{Xi_boundary} are equivalent to the problem of a point charge next to a conducting sphere. Solving these equations using the method of images yields
\begin{equation}
  \Xi(\bar{\pos}) \sim \frac{1}{|\bar{\pos} - \bar{\pos}_{0}|} - \frac{1 / \alpha}{
    |\bar{\pos} - \frac{1}{\alpha^{2}} \bar{\pos}_{0}|}.
\end{equation}
Upon switching to spherical coordinates $\bar{r} = |\bar{\pos}|$ and $\Omega = \cos(\theta)$, we have
\begin{equation}
\begin{split}
  \Xi(\bar{r}, \Omega) \sim \frac{1}{\sqrt{\bar{r}^{2} + \alpha^{2}  - 2 \alpha \bar{r} \Omega}}
  -  \\ 
  \frac{1}{\sqrt{1 + \alpha^{2} \bar{r}^{2} - 2 \alpha \bar{r} \Omega}}.
\end{split}
\end{equation}

We are now ready to calculate the first-hit distribution $q(\Omega)$. This distribution is proportional to the time integral of the radial flux on the sphere. First, note that the integrated radial flux in general $\lambda(\bar{r}, \Omega)$ is
\begin{equation}
  \lambda = \int_{0}^{\infty} \frac{\partial \xi}{\partial \bar{r}} d\tau =
  \frac{\partial}{\partial \bar{r}} \left( \int_{0}^{\infty} \xi d\tau \right)
  = \frac{\partial \Xi}{\partial \bar{r}}.
\end{equation}
After evaluating $\lambda$ on the sphere, $\bar{r} = 1$, and adding a normalizing constant, we have
\begin{equation}
 q(\Omega) = \frac{\alpha^{3} - \alpha}{2(\alpha^{2} - 2
    \alpha \Omega + 1)^{3/2}}.
\end{equation}
Lastly, to get the first-hit cumulative distribution function (CDF) $Q(\Omega)$, we integrate from $-1$ to $\Omega$
\begin{equation}\label{cdf}
  Q(\Omega) = \frac{(\alpha - 1) \left(\alpha + 1 - \sqrt{\alpha^{2} -
        2 \alpha \Omega +1} \right)}{2 \sqrt{\alpha^{2} - 2 \alpha \Omega
      + 1}}.
\end{equation}

Although Eq.~\eqref{cdf} is normalized, we have not yet addressed the transience of Brownian motion in three dimensions. Unlike in 2D, there is a chance the particle will escape to infinity before absorbing at the boundary. Specifically, this escape probability is \cite{wendelHittingSpheresBrownian1980}
\begin{equation}\label{escape}
  \mathrm{Probability(Escape) = 1 - \frac{1}{\alpha}}.
\end{equation}

As a result, sampling the 3D first-hit distribution requires two steps. First, we must make a draw against the escape probability in \eqref{escape} to see if the particle makes contact with the sphere at all. If it does, we sample \eqref{cdf}, the CDF, by setting $Q$ equal to a random number taken from $[0, 1]$ and solving for $\Omega$
\begin{equation}\label{inverted_cdf}
  \Omega = \frac{\alpha^{2} + 1}{2 \alpha} - \frac{(\alpha^{2} -
    1)^{2}}{2 \alpha (\alpha - 1 + 2 Q)^{2}}.
\end{equation}
The azimuthal angle can then be drawn randomly from the interval $[0, 2 \pi)$. If, instead, the particle escapes in the first step, it is taken to have reached the distant, surrounding bath. In this case, the particle is killed, and a new diffusive particle is launched. 

\bibliography{references}